\documentstyle[prl,aps,psfig,twocolumn]{revtex} 

\begin{document}
\twocolumn[\hsize\textwidth\columnwidth\hsize\csname
@twocolumnfalse\endcsname
\title{Discretized Diffusion Processes.}
\author{Stefano Ciliberti$^1$, Guido Caldarelli$^1$, Paolo De Los Rios$^2$, 
Luciano Pietronero$^1$ and Yi-Cheng Zhang$^2$.}
\address{$^1$INFM Sezione di ROMA1 Dip. Fisica, Universit\`a di Roma
``La Sapienza'' P.le A. Moro 2 00185 Roma, Italy.}
\address{$^2$Institut de Physique Th\'eorique, 
Universit\'e de Fribourg, CH-1700, Fribourg, Switzerland.}
\date{\today}
\maketitle
\begin{abstract}

We study the properties of the ``Rigid 
Laplacian'' operator, that is we consider solutions of the Laplacian 
equation in the presence of fixed truncation errors.
The dynamics of convergence to the correct analytical solution displays
the presence of a metastable set of numerical solutions, whose presence
can be related to granularity.
We provide some scaling analysis in order to determine the value of the 
exponents characterizing the process. 
We believe that this prototype model is also 
suitable to  provide an explanation of the widespread presence of 
power-law in social and economic system where information and decision 
diffuse, with errors and delay from agent to agent.
\end{abstract}
\pacs{05.40+j, 64.60Ak, 64.60Fr, 87.10+e}
]
\narrowtext

Most equations in science are continuous in value. While this is a good 
approximation to reality, natural processes often are discrete in 
microscopic details ({\it e.g.} atoms and molecules). 
We ask the following general questions: if the variables are subject to 
some small threshold such that no changes smaller than such a threshold are implemented,
how do the continuous equations behave? Can we properly recover the continuous limit
letting the threshold vanish? Are there new features emerging independent on the
threshold value (and therefore robust down to the continuous limit)?  

This problem is somewhat general and this is the reason
of its importance. If we consider in particular the class of 
self-organized critical systems then the question becomes crucial. 
We know that for many SOC systems, such as the BTW \cite{BTW} or the
Zhang model \cite{zhang}, threshold and discretisation play a dramatic role in order 
to avoid the diffusive behaviour and to attain the critical one. 
By considering the continuous limit of the equations describing the 
microscopic dynamics of these cellular automata and by not preserving a finite
threshold we get a trivial diffusive behaviour instead of the 
critical one\cite{rigidity}. 
 
In order to investigate this kind of subject, 
we study here one of the oldest and most important equations of physics: 
the Laplace equation $\nabla^2 \phi(x) = 0$. Its time dependent version,
describes the diffusion of heat and of particles, and the 
relaxation of incompressible, non-viscous fluids.
In imaginary-time it corresponds to the Schroedinger equation of
quantum mechanics for a free particle (the difference vanishes completely
if we are interested in stationary solutions/eigenstates).
Therefore, we do not overestimate its importance if we
consider it to be ubiquitous in physics. More recently,
it was also considered to play a major role in fractal growth processes such
as Diffusion Limited
Aggregation (DLA)\cite{DLA} and the Dielectric Breakdown Model (DBM)
\cite{DBM}. 

It is thus not a surprise that many methods have been
developed to solve it, at least numerically. In general, 
both time and space are discretized on a lattice. The iterative method
is the most commonly used method of solution, especially for
problems such as DLA and DBM where the boundary conditions change
in time. Essentially, the method resorts to iterating the equation
\begin{equation}
\phi(i,t+1) = \frac{1}{2d} \sum_{j \; n.n. \; i} \phi(j,t), 
\label{equilibrium}
\end{equation}
whose fixed point correspond to the discretized version of $\nabla^2 \phi(x) = 0$.
Numerically convergence to the fixed values $\phi(i)=\phi(i,t=\infty)$ has 
to be defined by some error: when two successive configurations 
differ by less than some small $\delta$ then the iteration stops.
Of course $\delta$ has to be greater than the machine precision, $\delta'$.
This means that when the field $\phi$ is of order of unity, 
it is impossible to change its value of quantities less than $\delta'$. 

Therefore it is of extreme importance to explore what happens if we
explicitely set a precision in the definition of $\phi$, that is, if we
say that we can change its value, according to (\ref{equilibrium}), only
by integer multiples of some chosen $\varepsilon$.
This corresponds to consider the fields $\tilde{\phi}$
\begin{equation}
\tilde{\phi}(i)= Int_{\varepsilon}\left[ \frac{1}{2d} \sum_{j \; n.n. \; i} 
\tilde{\phi}(j)\right]
\end{equation}
$Int_{\varepsilon}[x]$ corresponds to taking the integer multiple of 
$\varepsilon$ closest to $x$
(other possible definitions of the integer part, such as the lowest integer
closest to $x$, have been explored and have been shown to give the same
results). 
The relevance of this problem is not only related to the numerical
solution of Laplacian equation, but it comes also from real problems:
granular materials, for example, behave sometimes as fluids. As such, we can
expect them to obey some
of the laws typical of fluids (hence also to relax according to the time
dependent Laplacian equation), 
yet their intrinsic granularity forbids movements of 
quantities smaller than a single grain. 
Under this respect we refer to the properties of $\tilde{\phi}$ as 
properties of rigid diffusion processes.
Discrete-discrete Laplacian fields $\tilde{\phi}$ 
(discrete in space-time, and discrete in the field $\tilde{\phi}$) 
can also play a role in modeling of economic and social systems.
In particular, for the simplest case  of the Laplacian ``harmonic'' 
operator the field is computed by averaging over the field neighbours.
If this is viewed as a social system where opinions of players are formed
by consulting people around, we would like to introduce a  more 
realistic version, where decisions are considered 
only when a certain threshold of information is overcome and the diffusion
is therefore discretized.  
This effect known as ``rigidity'' in the literature\cite{rigidity} 
is indeed responsible of the onset of criticality\cite{Gabetal}.
Since Self-Organized Criticality is believed to play a role both in
economic and social system we suggest that our results can be viewed
as a way to describe the appearance of criticality in such environments.

In this Letter we explore the behavior of this Rigid Laplacian Model (RLM). 
As we shall see, non trivial properties emerge already in one dimension.

\begin{figure}
\centerline{\psfig{file=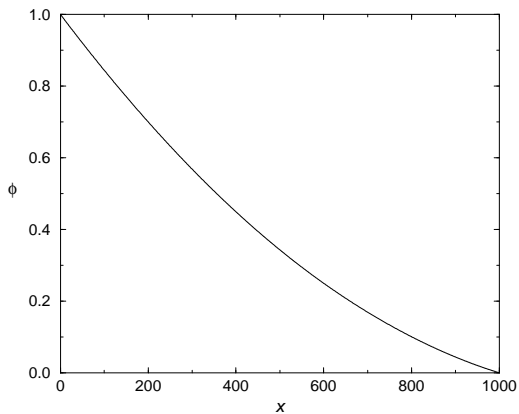,width=6.8cm}}
\caption{Profile of $\phi$ for a 
$d=1$ system whose length is $L=1000$. 
At boundaries $\phi(0)=1,\phi(L)=0$.}
\label{fig1}
\end{figure}

The standard setup of our simulations is the following: we take a $1d$ 
lattice of $L+1$ sites, and we set the boundary conditions $\phi(0) = 1$ 
and $\phi(L)=0$. Then we iterate the equation
\begin{equation}
\phi(i,t+1) = Int_{\varepsilon}\left[ \frac{1}{2} (\phi(i-1,t) + 
\phi(i+1,t)) \right]
\label{discdisclapl}
\end{equation}
The iteration stops when there are no more possible 
rearrangements. 
The exact solution of $\nabla^2 \phi =0$ on the lattice with 
the above boundary conditions is the
straight line $\phi(i) = -i/L + 1$. As we show in Fig.\ref{fig1},
the typical solution of (\ref{discdisclapl}) is instead a quadratic curve.

This can be easily explained by considering that as soon as a certain 
precision $\varepsilon$ is introduced, the Laplace equation can be considered as
a first approximation as
\begin{equation}
\nabla^2 \phi = \alpha \varepsilon
\end{equation}
In the case where the boundary condition are $\phi(L) = 0$ and 
$\phi(0) = 1$ we obtain the following equation 
\begin{equation}
\phi_{\varepsilon}(x) = \frac 12 \alpha \varepsilon  L^2 \left(\frac x L\right)
^2 - \left(1+\frac 12 
\alpha \varepsilon L^2\right) \frac x L +1,
\label{profilo}
\end{equation}
in terms of the reduced variables $x'=x/L$ and $\varepsilon'=\frac 1 
2\alpha \varepsilon L^2$ one obtains 
\begin{equation}
\phi_{\varepsilon'}(x') = \varepsilon'x'^2 - (1+\varepsilon') x' +1.
\end{equation}
The precise value of $\alpha$ depends on the initial condition and it has 
to be determined with some fitting procedure. 
We can study the dynamics of the system observing that
the evolution (\ref{equilibrium}) always minimizes the functional
\begin{equation}
E = \sum_{i=0}^{L-1} (\phi(i+1)-\phi(i))^2
\label{energy}
\end{equation}
Eq.(\ref{energy}) maps the RLM into the discrete space-time Gaussian model.
Physically, since $\phi$ represents an electrostatic field, $E$ represents
the total electrostatic energy of the sample, whose minimization in the
case of fractal growth has been recently studied\cite{MBCMR}.
Letting the field $\phi$ take on only integer values
leads to the {\it discrete} Gaussian model. The dynamics described by
(\ref{discdisclapl}) is therefore a $T=0$ dynamics, always decreasing the 
energy, and the stability of solutions different from the constant slope 
implies the presence of local energy minima.
Since this process of energy minimization stops in local minima, we can 
perform some sort of simulated annealing by perturbing it. 
We increase the value of the field $\phi$ in some randomly 
chosen point by $n\varepsilon$, then we let the system relax.
Indeed, we find that the curvature $\alpha$ decreases in time
(time is defined in terms of perturbation steps;
we consider the relaxation process to be much faster, as usual
in self-organized critical models).
Each perturbation implies a rearrangement of the profile. 

\begin{figure}
\centerline{\psfig{file=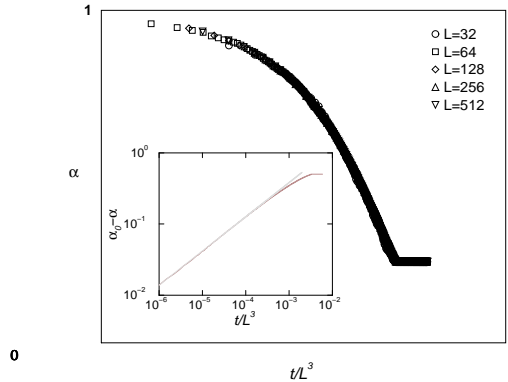,width=6.8cm}}
\caption{Plot in rescaled time of $\alpha(t)$ and $\alpha(t)-\alpha_0$}
\label{fig2}
\end{figure}

\begin{figure}
\centerline{\psfig{file=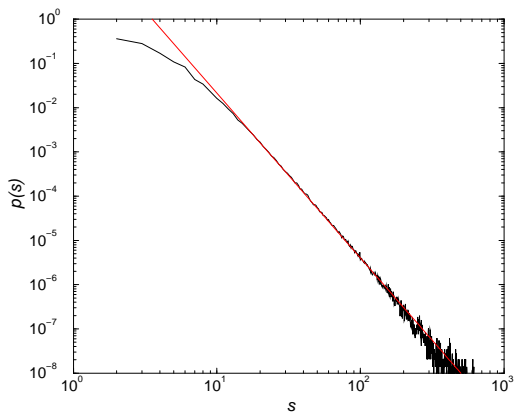,width=6.8cm}}
\caption{Plot of the density function $p(s)$ giving the number of times that
$s$ sites are updated in an avalanche.}
\label{fig3}
\end{figure}

For the energy of the system one has from the eq.(\ref{profilo}) 
\begin{equation}
E\propto L^3 \alpha^2 \varepsilon^2.
\label{ener}
\end{equation}
We know that for ordinary diffusion the energy variation $E$ is proportional to 
$t^{-3/2}$. We will show in the conclusion that based on that result,
{\em in this case}
one can expect $E\propto t^{1/2}$ or more precisely $E-E_0\propto (t/L^z)^{1/2}$, 
where we explicitely introduced a dynamical exponent $z$ for the process.
Assuming for the moment this result, one has 
\begin{equation}
E \! \propto \! E_0-(t/L^z)^{1/2} \! \Rightarrow \!(t/L^z)^{1/2} 
\!\propto \!E_0-L^3\alpha^2(t/L^z)
\end{equation}
By derivating this expression and by requiring that it shows no dependence on
$L$ (unless for the ratio $t/L^z$) one obtains that 
the dynamical exponent must be $z=3$, in agreement with the value 
$z=3.0\pm 0.1$ from numerical simulations, and
\begin{equation}
\alpha(t,L)=\alpha_0-(t/L^z)^{1/2} f(t/L^z).
\end{equation}
the scaling 
function $f(x)$ is such that $\alpha\rightarrow\alpha_\infty\neq 0$ for 
$t\gg L^z$. Such behaviour has been tested in Fig.\ref{fig2}
We see that for this kind of systems the relaxation process to 
the equilibrium is very slow because of granularity. Indeed, for a 
continuous Laplacian field the relaxation to the stationary state 
is an exponential 
process with a characteristic time scale $t_0\sim L^2$.
Such rearrangements proceed through power-law distributed avalanches, 
whose measure is the number of sites that changed their value of $\phi$.
We show in Fig.\ref{fig3} the probability $p(s)$ to deal 
with an avalanche of size $s$. This function can be fitted with a power
law of the kind $p(s) \propto s^{-\tau}$.
The value of the exponent $\tau$ measured from the data is $3.8 \pm 0.3$.
To take in account the finite size effects of this distribution 
(the size of an avalanche cannot be greater than the size of the system) 
we can write the universal form $p(s,L)=s^{-\tau}\,g(s/L)$, where 
$g(x)$ is a scaling function such that $p(s=L)=0$. 
This scale invariant behavior describes the dynamics in a transient period 
where perturbations are accumulated into the system, until a limit profile 
(not necessarily the straight one) is reached. 
This transient state can be suitably delayed by considering
a different dynamics where also negative perturbations are considered.
To describe the evolution to the stationary state, we 
measured the average size $\langle s \rangle$ of the 
avalanches with respect to the time of the simulation.

\begin{figure}
\centerline{\psfig{file=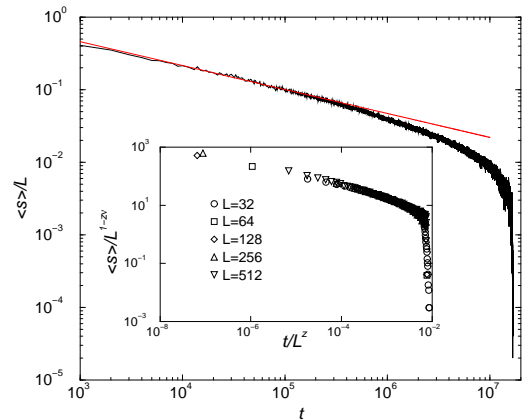,width=6.8cm}}
\caption{$\langle s(t) \rangle$ versus t, in the inset collapse
 plot for different $L$'s.}
\label{fig4}
\end{figure}

In Fig.\ref{fig4} we present the behaviour of $\langle s(t)\rangle$ 
that also behaves as a power law with exponent $-1/3$. 
A simple scaling argument can be presented
to explain such behaviour. Indeed,
\begin{equation}
\langle s (t,L)\rangle = L\,t^{-\nu}g(t/L^z)
\end{equation}
where we explicitely considered the finite size effects in the 
average size. In the above expression, $\nu$ represents the expected exponent
and $z$ is the dynamical exponent previously introduced and present also 
in the scaling function $g$.

The collapse shown in the inset of Fig.\ref{fig4} has been obtained with the
following values: $z=3.0 \pm 0.1$, $\nu=0.34 \pm 0.01$.
The values of these two exponents are related, indeed
the meaning of the dynamical exponent is as usual to determine the scale of
time after which the process stops. 
By imposing that at the maximum time the average size of the avalanche is
independent on $L$, one obtains the scaling relation
\begin{equation}
1-z\nu =0
\end{equation} 
from which $\nu=1/z=1/3$ as numerically found in the limit $t << L^3$.

From this scaling relation it's possible to provide an argument to
determine the value of the exponent $\tau$.
Since the number of avalanches is the same whether they are classified 
according to their sizes (i.e. by means of the $p(s)ds$) or just counted
in time (i.e. by means of $dt$) then one obtains 
$p(s)=\frac{dt}{ds}$.
From the above scaling laws one finally obtains 
\begin{equation}
\tau=1+1/\nu.
\end{equation}

\begin{figure}
\centerline{\psfig{file=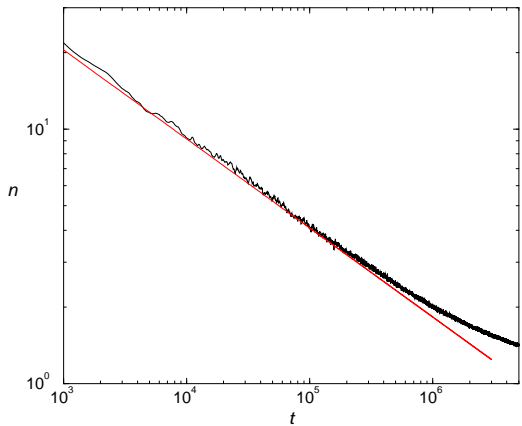,width=6.8cm}}
\caption{Plot of $n(t)$; the dashed line is $y=t^{1/3}$.} 
\label{fig5}
\end{figure}

We numerically tested the validity of these relationships as above mentioned.
The ``rigidity'' of the field makes the process sub-diffusive as it is 
implied by the value $z=3$ of the dynamical exponent, as opposed to the usual 
$z=2$ value. A more intriguing feature of these processes can be 
obtained by a suitable rescaling of the microscopic time scale.
As a matter of fact one can check the number $n(t)$ of microscopical 
sweeps needed in order to update the values of the field $\phi$ in an 
avalanche.
This number $n(t)$ must be proportional to the average size 
of an avalanche at time $t$ and then it varies as $t^{-1/3}$, 
as it can be shown by a direct measurement (Fig.\ref{fig5}). 
It means that the ``physical'' time of
the process defined as $\sum_{t'=1}^{t} n(t')$ behaves as $t^{2/3}$.
By rescaling the scale of time with this characteristic time scale, the
process loses any information on the granular nature of the medium and
the dynamics behaves as a usual diffusive process.
In this way we demonstrate the initial assumption for the energy $E$ in the
system. In fact, since it varies for ordinary diffusion phenomena as 
$E \propto t'^{-3/2}$ and in this case the time of the process $t'$ varies
as $t' \propto t^{-1/3}$ we obtain the above assumption of $E \propto t^{1/2}$.

We also checked the properties of this model in dimension larger than $d=1$.
We found a similar scale invariant behaviour.
In the case of $d=2$ we found a value of $\tau=2.65\pm 0.03$, for $d=3$
we have $\tau=2.60 \pm 0.05$ and for $d=4$ $\tau=2.55 \pm 0.05$.

As described in the introduction, it would be tempting then to link 
this randomness to the known effect of ``rigidity'' in Self-Organized 
systems\cite{rigidity,Gabetal}. 
Self-Organization has often been invoked as a way to describe
the invariance observed in various social and economic 
systems\cite{zipf,mand,cald,ric}. 
Indeed we propose the basic mechanism of diffusion of money and/or
information in a random environment (that can stop or reduce the information
interchanged between players) as one of the reasons that could lead to  
scale invariance and to avalanche dynamics. 
As regards the economic systems, we believe that this study could be used also
to describe the time behaviour of a quantized field with respect 
to the value of the truncation error. 
A specific example in this field is be presented by considering the 
spurious effect in the price dynamics that could happen by changing the tick at
which they are traded (i.e. from $1/16$ of dollar to $1/100$).

In conclusion, we introduced a model of diffusion, where the redistribution 
process
is hindered by some sort of rigidity induced by the granularity of the fields.
Randomness is present through a cutoff in the 
different possible values the field can assume. 
By truncating to a certain precision the value of the field a new
and rich scenario of metastable states appears.
The evolution of this system through the metastable states is driven by an 
avalanche dynamics with no particular time or length scale.
By suitably rescaling the microscopic time of the evolution, one can use
the scaling relations of diffusive process to give a theoretical ansatz for
the quantities describing the system. From computer simulations we find a good
agreement between numerical data and theoretical ones. 
We believe this sort of process could be responsible for the 
ubiquitous presence of 
power-law relations in the every day life.
Clearly our work is readily generalised to other systems, 
notably Naviers-Stokes, KPZ\cite{marsi}, wave equations. 
One may wonder if there is some sort of super-universality
in their scaling behavior. Indeed a long list of similar problems awaits us to
study.

P. De Los Rios and Y-C. Zhang thank the Dipartimento di Fisica 
in the University of Rome ``La Sapienza'', for its kind hospitality.
G. Caldarelli thanks for the same reason the 
Institut de Physique Th\'eorique in Fribourg.
This work has been supported by the European Network contract FMRXCT980183.

\end{document}